# Development of a Miniaturized Deformable Mirror Controller

Eduardo Bendek*[a,b], Dana Lynch[a], Eugene Pluzhnik[a], Ruslan Belikov[a], Benjamin Klamm[a], Elizabeth Hyde[a], and Katherine Mumm[d]

[a]NASA Ames Research Center, Moffett Field, CA, USA 94035; [b]Bay Area Environmental Research Institute, Petaluma, CA, [d]Los Altos High School, Los Altos, CA

## ABSTRACT

High-Performance Adaptive Optics systems are rapidly spreading as useful applications in the fields of astronomy, ophthalmology, and telecommunications. This technology is critical to enable coronagraphic direct imaging of exoplanets utilized in ground-based telescopes and future space missions such as WFIRST, EXO-C, HabEx, and LUVOIR. We have developed a miniaturized Deformable Mirror controller to enable active optics on small space imaging mission. The system is based on the Boston Micromachines Corporation Kilo-DM, which is one of the most widespread DMs on the market. The system has three main components: The Deformable Mirror, the Driving Electronics, and the Mechanical and Heat management. The system is designed to be extremely compact and have low-power consumption to enable its use not only on exoplanet missions, but also in a wide-range of applications that require precision optical systems, such as direct line-of-sight laser communications, and guidance systems. The controller is capable of handling 1,024 actuators with 220V maximum dynamic range, 16bit resolution, and 14bit accuracy, and operating at up to 1kHz frequency. The system fits in a 10x10x5cm volume, weighs less than 0.5kg, and consumes less than 8W. We have developed a turnkey solution reducing the risk for currently planned as well as future missions, lowering their cost by significantly reducing volume, weight and power consumption of the wavefront control hardware.

## 1. INTRODUCTION

### 1.1 Adaptive optics applications for exoplanet detection

Since the discovery of the first exoplanet Pegasi 51b in 1995, more than 5,600 candidates have been detected and 1,900 planets have been confirmed. Currently, direct imaging detections account for a very small fraction of the planets that have been found. Until the Kepler NASA mission delivered its first results in 2010, stellar Radial Velocity (RV) performed by ground-based telescope surveys was the exoplanet detection workhorse. After 2010, the data captured by the Kepler mission enabled the detection of thousand of planets surpassing the RV exoplanet yields. Currently, there are multiple efforts in the astronomical community to obtain reliable exoplanet occurrence rates based on the data available[1]. Although the error bars are large, data shows 50% of the G and K stars can harbor a terrestrial planet, defined as 0.5 to 2 times the earth mass and within the star's Habitable Zone (HZ). This result and the lack of information about the planet mass naturally sets exoplanet direct imaging as the next step in advancing our knowledge about the next frontier.

We find the same recommendation in multiple NASA strategic documents. More specifically, the long-term NASA 30-year roadmap[2] clearly states the importance of direct imaging: *"Directly image the planets around nearby stars and search their atmospheres for signs of habitability, and perhaps even life."* Moreover, the Decadal survey breaks down the grand visions of the 30-

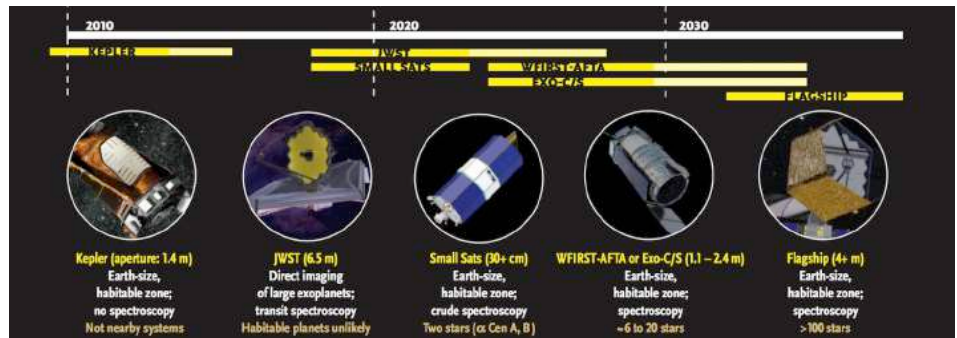

Fig. 1: A sampling of current and future space missions following Kepler that have coronagraphic direct imaging capability.

year roadmap further; stating that *"the plan for the coming decade is to perform the necessary target reconnaissance surveys to inform next-generation mission designs while simultaneously completing the technology development to bring the goals within reach."* (Astrophysics 2010 decadal survey[3]). The community has embraced these recommendations and has proposed multiple direct imaging missions focused on nearby stars, such as WFIRST[4], EXO-C/S[5,6], and a future flagship called HabEx. The timeline associated with these missions is shown in Fig. 1.

Direct imaging exoplanet missions must be able to achieve better than $10^{-9}$ contrast at fractions of arcsecond angular separations to be able to observe an earth-like planet around a sun-like star. These are challenging requirements currently beyond the performance of state-of-the-art optical systems. Multiple Starlight Suppression Systems (SSS) have been proposed and developed to meet direct imaging stringent requirements. Most of these systems are undergoing a testing phase in the lab as well as in ground-based telescopes. Internal coronagraphs, which are a family of SSS, require controlling the wavefront error level down to 1/10,000 of the observation wavelength to avoid speckles that will have a larger signal than the planet. Building, integrating and maintaining in orbit such a precise optical system would be extremely expensive. For example, if standard high-quality optics and a modern coronagraph such the Phase Induced Amplitude Apodization[7] (PIAA) are used, the raw contrast is limited to $1 \times 10^{-5}$ raw contrasts due to optics aberrations and alignment errors.

To overcome this challenge current direct imaging exoplanet testbed demonstrators utilize a Deformable Mirror (DM) to correct wavefront errors caused during manufacturing and/or alignment. Deformable mirrors have allowed different facilities to reach record contrast levels of $1 \times 10^{-8}$ in air at the NASA Ames Coronagraphic Experiment laboratory (ACE) and $1 \times 10^{-9}$ in vacuum at JPL. Over the years many different wavefront control algorithms have been proposed and tested to create dark zones controlling a DM, such as, Speckle Nulling, Electric Field Conjugation[8] (EFC), and Stroke Minimization among others. These have been used at the ACE[9], Princeton, JPL and Goddard. There are also emerging technologies such as the Multi-Star Wavefront Control (MSWC)[9], which uses the DM to control the light dispersed by aberration in case of binary stars. This technology is being implemented using a PIAA coronagraph by the Ames Coronagraph Experiment (ACE) team who has been pioneering and maturing the technology[11].

All the internal coronagraph based missions require a Deformable Mirrors to deliver their design performance. However the deformable mirror controller technology is not ready, therefore, it has been identified for three consecutive years, as a Coronagraph Technology Gap[12] by the NASA's Exoplanet Exploration Program (ExEP).

## 1.2 Deformable mirror technologies

There are two main DM actuator technologies currently being considered for space missions; the first one is the electrostrictive actuator manufactured by Northrop Grumman Xinetics, and the second one is the electrostatic force Micro Electro Mechanical System (MEMS) DM manufactured by Boston Micromachines Corporation (BMC). The driver electronics for both technologies are challenging. They require controlling high-voltages at high speed, however, the electrostrictive actuators act as a capacitor, therefore every time an actuator is moved there is a "non-negligible" current. In contrast, the MEMS DM working with electrostatic force only requires a voltage at the actuator pin, so there is no current associated with an actuator motion. This facilitates the high-speed control electronics necessary for many missions. This difference becomes more important as the number of actuators grows. The ExEP envisions 128x128 actuators DM for future 10m flagship missions. Such DMs would have more than 16,000 and therefore will present more difficult challenge to controlling current and voltage.

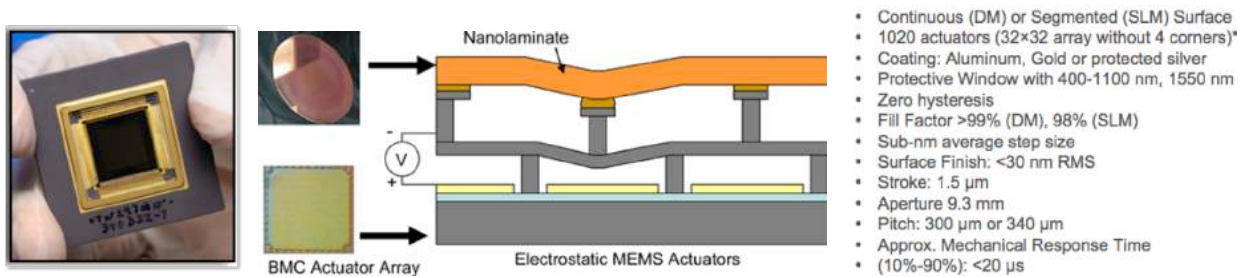

Fig. 2: BMC Kilo DM deformable mirror and its typical characteristics

The BMC Kilo DM has an array of 32x32 actuators each one with 300μm X 300μm pitch each. The electrostatic force caused by a potential difference between the plate below the actuator and the actuator support moves each actuator. The maximum deflection is in the order of 1.5μm when ~250V are applied. The exact values changes from DM to DM, since this is a non-linear force, the voltages v/s deflection are very sensitive to the distance between the actuator and the bottom plate, which is different for each device. Figure 2 shows the DM, a construction schematic and its specifications.

There are three desirable qualities for a deformable mirror: 1) the stroke, that allows to correct larger aberration; 2) the number of actuators that correlates with Outer Working Angles (OWA) or corrected Field of View; and 3) the resolution that defines how accurate the DM can control the wavefront and therefore the limiting contrast of the SSS. Modern coronagraphs utilize at least 1,000 actuators with strokes of ~1.2μm which results in 200V potentials to be applied and a resolution of at least 14bits over the voltage dynamic range. These requirements result in bulky, heavy and high power consumption DM controllers that do not present a major issue for laboratory development. However, they are a showstopper to enable the DM application on space missions, especially on small ones. Currently the most widespread DM is the Kilo DM manufactured by Boston Micromachines. The controller utilizes a 19" Chassis (5.25"x19"x14") that consumes up to 40W and it weights almost 10kg.

### 1.3 Objectives

The objective of the proposed work is to *facilitate the utilization of MEMS Deformable Mirrors in space by developing a space-capable and miniaturized deformable mirror controller* for exoplanet imaging and a wide range of space missions with optical applications, such as direct line of sight laser communication. This work reduces cost, risk and development time for those missions. We decided to develop the controller for the Boston Micromachines Kilo DM because it is suitable for space applications and it is being utilized in many, laboratories, such as NASA Ames, Princeton University, JPL and others as wells as on ground-based exoplanet detection instrument such GPI at Gemini South and SCxEAO at Subaru telescope. Our goals are the following:

- **Goal 1: Develop a miniaturized controller** for the BMC Kilo DM that can fit on a 10x10x10cm with a dynamic range of 1μm and with 14bits effective precision and 16bit resolution.
- **Goal 2: Advance the state of the art** towards future exoplanet mission challenges including:
  - **Prove a scalable DM Controller architecture** that can be scaled in preparation for large actuator number DMs 64x64 or 128x128.

These goals are aligned with NASA's strategic documents and the ExEP, and from both of them we obtain our top-level and derived requirements shown in table 1.

Table 1: Deformable mirror controller requirements traceability matrix

| Top-level requirement | Rationale | Derived requirement | Value | Implementation value |
|---|---|---|---|---|
| Create a dark zone at out to from 1.6 to 12λ/D | Sample the HZ and beyond the ice-line of G, H and K stars within 5pc with 1m telescope or smaller | Have more than 24 actuators across | Kilo DM with 32x32 actuators. | 32x32 actuators |
| DM stroke and resolution to enable raw contrast of <$10^{-9}$ at 20% bandwidth | Imaging earth like planets will require $10^{-9}$ raw contrast | At least 1μm stroke | 1.2μm | DE with 200V dynamic range |
| | | Actuator resolution of < 10 Armstrong | 60 Armstrong Resolution | 14bit precision, 16bit resolution voltage control |
| | | DM position stability < 0.1μm | 0.1μm thermal expansion | Invar mechanical connection directly to the Zif socket. |
| Wavefront and jitter control | Use one DM for both functions | 1kHz max operational speed | 1kHz refresh rate | Real Time Computer and High-Voltage amplifiers |
| Space-qualified | To enable space missions use | Low Earth Orbit Rad levels for 2 years | TID > 30krad | SEU Th. LET: 20 MeV/mg/s^2  SEU Error Rate: $10^{-7}$ err/bit-day |
| Cubesat compatible | Enable other uses, parts commonality with large missions | Mass <1 kg | 0.5kg | 0.5kg |
| | | Volume <1U | 10x10x10cm | 10x10x5cm |
| | | Power less <10W | 8W max | 8W max |

## 2. SYSTEM ARCHITECTURE AND DESIGN

### 2.1 System architecture and subsystems

We developed a system to accomplish the goals and specified in section 1.3. The system has three main functionalities: The DM, which is an off-the-shelf component provided by BMC, the Driving Electronics and Real Time Computer (DE & RTC), and the Mechanical and Heat Management (MHM). The functional block diagram is shown in Fig. 3. The DE provides the communications with the instrument or control computer using a USB 3 interface, which transmits the new position of the actuators using 32bits 32x32 FITS files at 1kHz requiring a data rate of 4.2MB/s. Each pixel value represents the desired voltage on the actuator. This standard is very common on different test beds. In addition, there is a telemetry communication channel providing information about the actual voltages applied, system temperature and power consumption. 12V power is supplied with maximum consumption of 8W. The MHM provides mechanical support as well as heat sink for the DE. The MHM connects with the instrument with an array of four #8-32 threads arranged on a 1-inch side square. Also the MHM holds the DM Zero Insertion Force (Zif) socket and connects it directly to the instrument interface with invar bars to avoid using DM stroke to compensate for thermal expansion. The DM receives the command voltages directly from the DE, however its position is determined by the MHM.

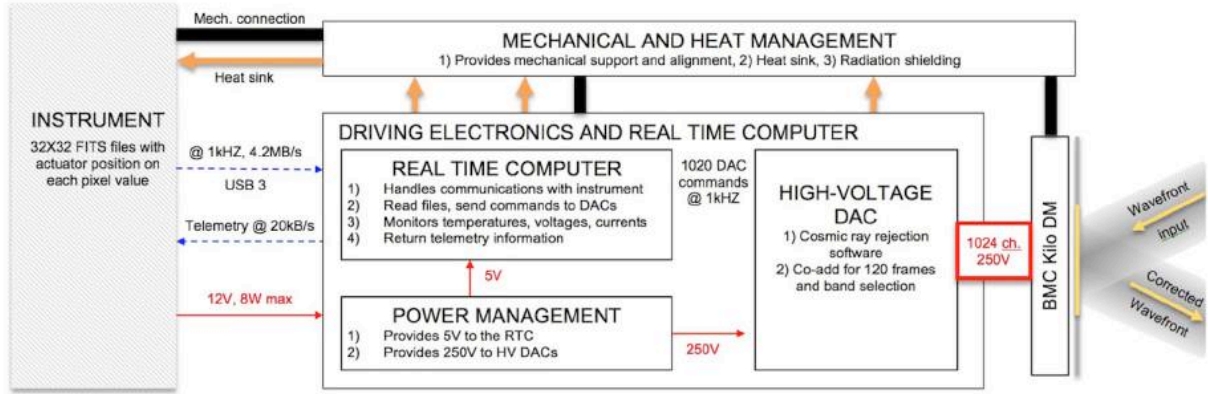

Fig. 3: Deformable Mirror controller system architecture. There are three main subsystems: The DM, Driving electronics, and Mechanical and Heat Management. Their interactions are shown above.

### 2.2 Deformable Mirror

For this project we selected BMC Kilo DM. Our lab at NASA Ames has access to engineering unit that has a fraction of the mirror surface damaged. However its functionality is not compromised on areas where the mirror is intact. This unit allowed performing initial testing without the risk of damaging a science grade unit. After the safe operation of the DM was proven a fully functioning Kilo DM was installed for testing. The units are shown in figure 4.

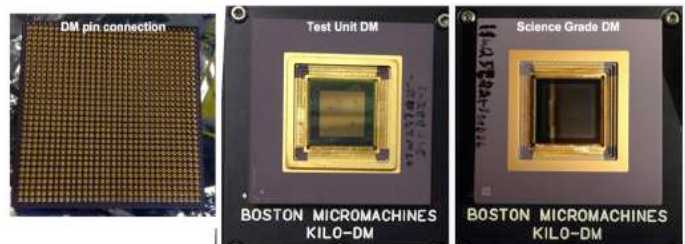

Fig. 4: Kilo DM deformable mirror controller units used for testing. Left is the pin array; Center is the test DM used the first controller test; Right is the science-grade unit used for characterization.

The Kilo DM has 1024 pins, and 1020 actuators. The pins on the four corners are common grounds. There is a mapping that relates the position of pins on the back of the chip and the DM actuator. This mapping has been provided by BMC and it is different for the square and rounds versions of the Kilo DM

### 2.3 Driving Electronics and RTC

The core complexity of the controller is to generate a large number of high-voltage channels ranging from 0 to 220V for full stroke DM motions. There is no ASIC semiconductor chip available in the market that can convert 1024 channels transmitting 16-bits digital TTL signals into 0 - 220V analog signals. To solve this approach we integrate multiple smaller ASICs that are available in the market. This is similar to the Graphical Processing Units (GPU) boards have taken to deliver unprecedented low-cost computing processing power.

Our design is based on an array of multi-channel high-voltage sample and hold devices (HV-DAC), such as Dalsa DH9685A, which has 96 channels. The real time computer will be implemented using a PIC MZ Series microprocessor, which receive the commands from the instrument/computer and sends a digital signal to the HV-DACs. Finally, a high-voltage power supply was built in to provide a maximum of 240V from a 12V power supply. Fig.

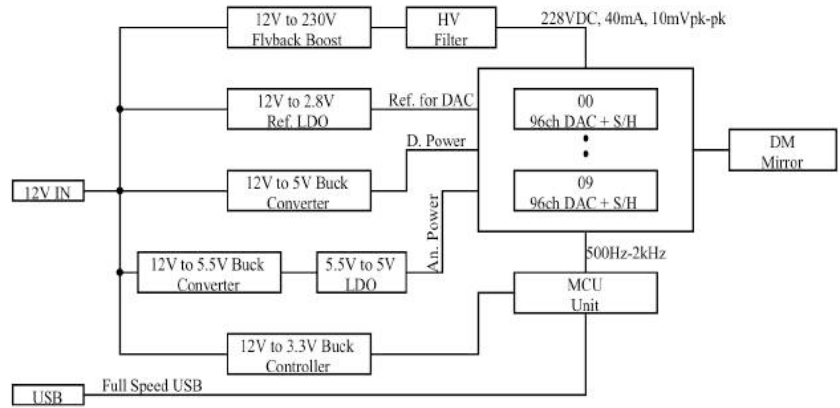

Fig. 5: Basic DM Controller Architecture

5 shows and schematic of the controller architecture. The electronic design and PCB layout considers the difficulties related to the integration of high-output impedance analog drivers with high-speed digital electronics and switching power supplies.

The design has a 12-Layer Flex-Rigid PCB with a soldered ZIF socket. Final footprint is subject to further evaluation, but initial design is shown on figure 7. Utilization of flex-rigid PCB allows for higher reliability due to the lack of high-density connectors and a smaller footprint. Teilch as a subcontract will develop the HV-DAC boards and the Microprocessor units.

**2.4 Mechanical And Heat Management**

The system has been designed to provide a stable and rigid mechanical interface with the DM. The Kilo DM is mounted on the Zero Insertion Force (Zif) socket custom made for BMC. The Zif socket is mechanically attached to an invar plate, which has 4 legs going around the driving electronics as shown in Figure 6. The legs are mounted on an Aluminum standard mounting plate to match CTE with the typical Aluminum mounts found in these systems. The legs are made of invar to avoid that localized heat source expand a leg and creates a tilt in the DM and therefore reducing it effective stroke. The legs are compliant enough in the lateral direction to absorb the connection aluminum plate expansion. Note that the ZiF socket is mechanically over constrained by the invar legs and the pins connecting with the PCB. However, the system has been designed to allow the pins connection to be more compliant than the invar mounting system. The high-voltage power supply and each of the HV-DAC will be dissipating most of the power consumed by the controller. In the absence of convection, heat sinks that can remove the heat by conduction are necessary. We will add metallic heat conductors on top of every HV-DAC and the high-voltage power supply. Those heat sinks are connected to a central heat pipe connected to the instrument.

**2.5 Interfaces and Control Software**

The controller requires only two connections: Data and power. The data is sent to the DM controller through a USB 3 connection. The DM commands are transmitted as 32bit FITS 32x32 pixel images. The value of each pixel represents the voltage to be applied to the respective actuator. The controller can run up 1 kHz, which requiring a data rate of 4.2MB/s. The USB interface also allows transmitting real-time telemetry from the controller at 20kB/s. This interface is shown in Figure 3. Power is supplied using a standard 12V connector. A PIC microprocessor running in real-time, reads the images and the digital values to the HV-DAC for analogic conversion and high-voltage amplification. The microprocessor also control the high-voltage power supply and reads telemetry.

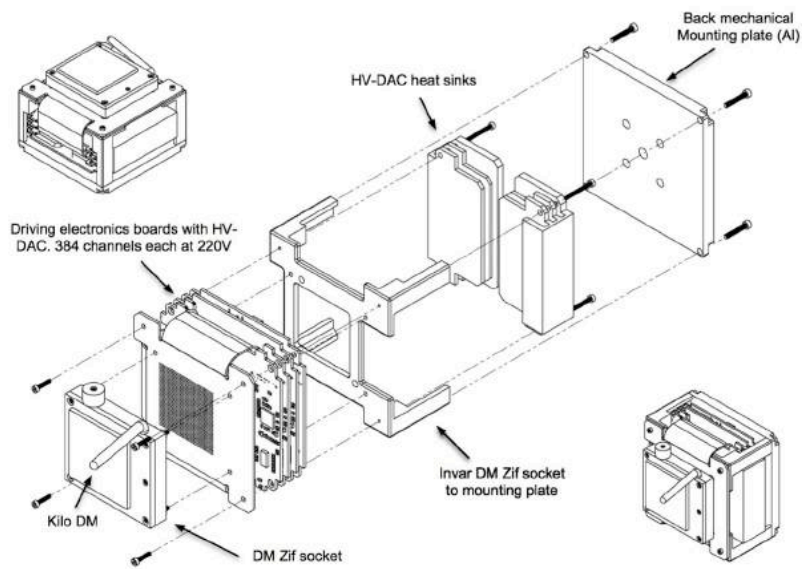

Fig. 6: DM controller assembly model including electronics and mechanics

On the user side, Microsoft Windows control software, shown in Fig. 8, has been developed to control the system from a computer. The software finds a COM port to connect. Once the system is connected, it is possible to browse files to select the desired FITS file. Also, if the continuous update check box is selected, the system will continuously read the selected file at 1kHz frequency. The control software overwrites the file to send a new DM command. A flag to make sure the file write was completed was implemented to avoid conflicts with the reading process.

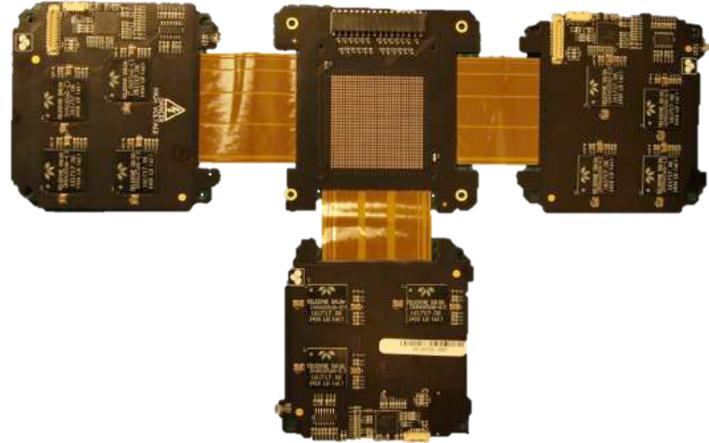

Fig. 7: Flex-rigid PCB used to mount the HV-DACs.

## 2.6 Space capability design considerations

This controller has been designed to evolve into a "Space qualified" deformable mirror controller. Here we outline our plan to evolve into a space-qualified version of the current controller. There are established guidelines for general mission planning purposes. The guiding document for Class-D missions at Ames Research Center is APR 8070.2 "Class D Spacecraft Design and Environmental Test". This document pulls largely from GSFC-STD-7000A "General Environmental Verification Standard (GEVS)" and JPL Design Principles (Rev. 3). When APR 8070.2 does not provide enough detail, GEVS test levels and methodology are used. While most Class-D projects use a proto-flight testing philosophy, this component will be tested at qualification levels.

**Radiation hardening considerations**
The system can be hardened against measurable Total Integrated Dose (TID) effects up to 30 krad, and will integrate options for direct package shielding to increase the service lifetime for mission-specific radiation environments (i.e.- electron-focused shielding for GEO and hadron-focused shielding for L2 and LEO). The system will be robust against single event effects through the use error detection and correction software (EDACS), Single Event Latch-up (SEL) immune power switches, active power filtering for voltage spike transients and single event burnout (SEB) prevention, and a watchdog timer will hard reset the system in the event multiple Single Event Upset (SEU) overload the EDACS. The conformal coating will be electrostatically grounded to mitigate deep di-electric charging. To mitigate the risk of unpredicted behavior of high-voltage components under radiation effects we will develop and test two different radiation-hardening architectures. We will manufacture two 32-channel systems for validation of **Architecture A** and two 32-channel systems for validation of **Architecture B.** We will choose the most appropriate one for the controller development after testing both under real radiation environment. The shield material to be used for the package is a HMW phenolic novolac doped 35%-by-weight with a tungsten powder and has aluminum foil surface passivation layers to ground any di-electric charging that may occur.

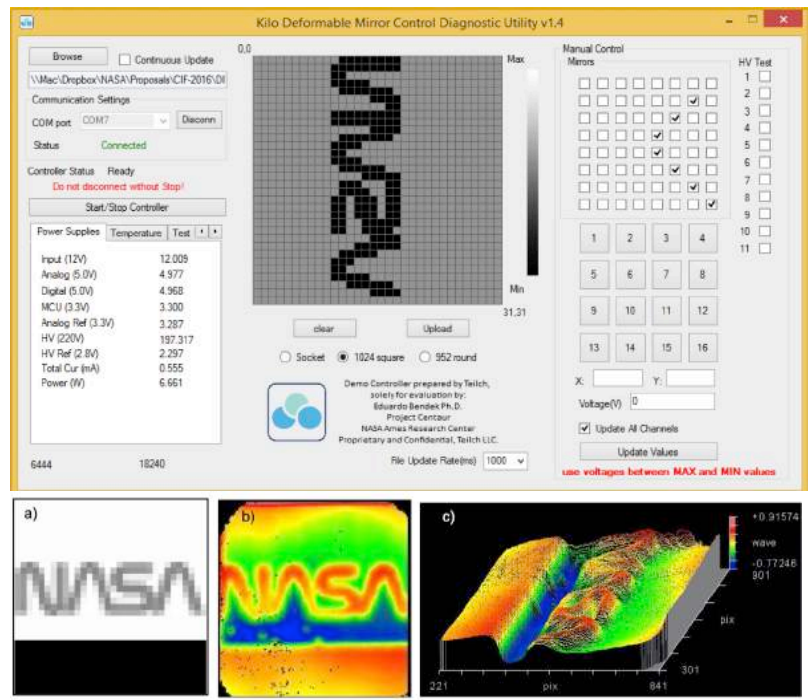

Fig. 8. Top: DM Controller software interface. On the left there is the FITS file browser, communications, and high-voltage amplifier switch. Below those we find the Power and Temperature telemetry. On the center the voltage to be applied are shown, on the right there is manual actuator control. Fig. 7a: FITS file image 32x32 pixels in 32 bits depth. White is 150V and black is 0V. Fig. 7b: Interferogram of the mirror surface after applying the file. Fig. 7c: Same as 7b but on an isometric view.

**Architecture A**
- Utilization of shield specified decreasing the radiation levels to meet the 5krad does established by NASA Reliability Practice No PD-1258 for commercial components.
- Distributed microcontroller (MCU) architecture to identify and correct SEU error rates.
- Typical components: MCU: PIC MZ Series, HV-DAC: Dalsa DH9685A

**Architecture B**
- Utilization of Rad Hard components with additional ultra-high molecular weight polyethylene shielding to provide better radiation resilience.
- Distributed microcontroller (MCU) architecture to identify and correct SEU error rates.
- Common low voltage DAC and S/H and independent high voltage channels.
- Typical components: MCU: UT80CRH196KDS, LV-DAC Sample and Hold: RHD5922, Op Amp: RH1078M

**Vacuum and high voltage:** MEMS devices can be damaged by corona discharge or electrical breakdown. The voltage to trigger the discharge is a function of the pressure and gap length; this equation is known as Paschen's law. The lowest voltage that will trigger a corona discharge is 250V between conductors separated by 10mm and 130Pa ambient pressure of predominantly Nitrogen atmosphere, which will occur during when the launcher reaches more than 33km. However, the discharge voltage increases rapidly for pressures of less than 10Pa. Therefore at LEO, where the pressure is $10^{-8}$ Pa, there is no risk of corona discharge. To avoid any damage to the system by accidentally powering the unit during launch, a pressure sensor interlock will be installed in the high-voltage power supply preventing high-voltages during the corona pressure regime.

**Outgassing protection:** Component outgassing can condensate in the Deformable Mirror surface affecting the mirror reflectivity. Therefore, all flexible conductors that can outgas will be isolated with Keptan. All components will be cleaned assembled and coated with PCB conformal coating to avoid outgassing.

**Heat management:** The high-voltage power supply and each of the HV-DAC dissipates most of the power consumed by the controller. In the absence of convection, heat sinks that can remove the heat by conduction are necessary. We will add metallic heat conductors on top of every HV-DAC, as well as the high-voltage power supply. Those heat sinks will be connected to a central heat pipe attached to the instrument.

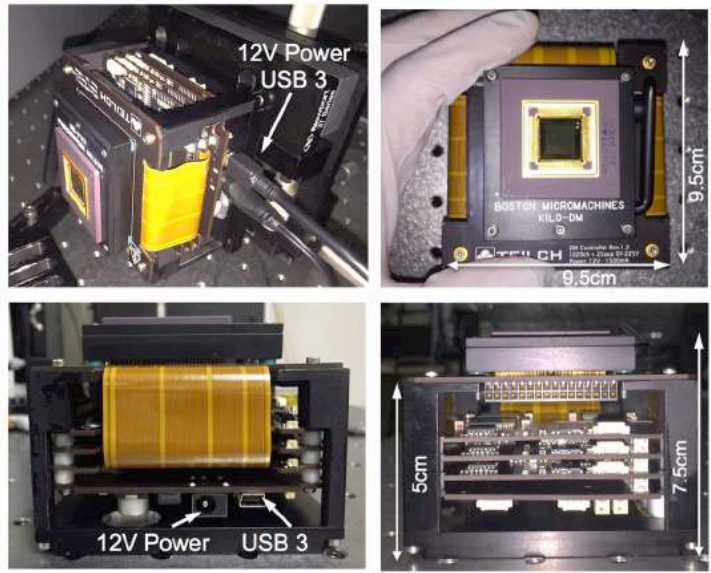

Fig. 9: Images of the DM controller used for testing after mounting the DM on the ZIF socket, connecting 12V power and USB connection (upper-left).

**Shock and vibration:** The system, including the mount, enclosure, components soldering, and Zif socket, will be designed to pass the shock and vibration tests

# 3. QUALIFICATION AND TESTING

## 3.1 Prototype description

We ordered the manufacturing of the electronic boards from an external company that implemented the architecture described in section 2.3. We integrated the boards with the mechanical structure and the ZIF socket was attached to the central board to which the flex PCBs are connected. Each board can control 396 actuators, so it is possible to stack more boards to control DMs with larger actuator counts. The assembly process is short and simple, enabling a large number production if necessary. The USB and 12V power connectors are directly accessible on the left side of the controller. The DM controller is 9.5 x 9.5 x 5cm and therefore fits on a 0.5U volume and form factor. The system weight less than 0.5kg and consumes less than 8W, therefore complying with the requirements specified on Table 1. The back plate has a standard ¼"x20 screw pattern with 1" spacing, allowing to attach the controller and its DM to any standard optical bench. Upon received the controller we unpacked and attached it to the ACE Laboratory optical bench. Afterwards, we connected the controller to the 12V power supply and USB data to the computer to begin the electronics test. The system is shown in Figure 9.

## 3.2 Electronics testing

The MEMS DMs operation is based on electrostatic force caused by a potential difference, and no current circulation. Therefore the voltage measurements in the ZIF pins should be representative of the voltages with the DM installed because it does not induce any load. An oscilloscope was connected to a randomly selected pin and to a ground pin located on each corner. Also, the controller provides real-time telemetry about components temperature and power consumption. Table 2 shows a summary of the final electronics configuration, power consumptions and voltages on the main components after operating for 1hr at 160V.

Table 2: Summary of the electronic testing and voltage ramps

| Electronics | | |
|---|---|---|
| Available channels | 1056 (96x11) | 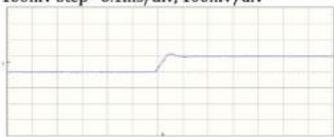 |
| Active channels | 960 (one HV-DAC was damaged, to be replaced) | |
| Dynamic Range | 180V (up to 225V adjusting HV power supply) | |
| Resolution | 16 bit = 2.7mV | |
| Accuracy | 14bits = 10.9mV |  |
| Power consumption | 6.6W @ 12V | |
| Max refresh rate | 1 kHz (Upgradable to 3kHz) | |
| HV-DAC refresh | 3kHz to achieve <1LSB stability | |
| **Power and Voltages after 1hr of operation** | | |
| Input 12V | 12.009 | 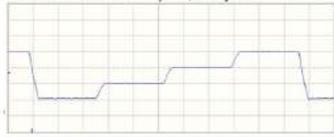 |
| Analog (5.0V) | 4.997 | |
| Digital (5.0V) | 4.968 | |
| High Voltage Supply | 197.317 (To allow max control 180V) | |
| Total current [mA] | 0.555 | |
| Total power [W] | 6.661 | 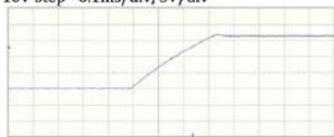 |
| **Subsystems temperature after 1hr of operation, no heat sink [˚C]** | | |
| Ambient | 19 | |
| ZIF socket internal | 29.983 | |
| ZIF socket external | 29.087 | 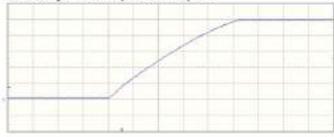 |
| High Voltage DACs | Range between [49.835 HV3 to 60.588 HV11] | |
| Main PCB | 44.703 | |
| Top board | 46.790 | |
| Middle board | 50.500 | |
| Bottom board | 52.110 | |

Note that the High Voltage Supply increases the voltage from 12V to 197V to enable a 1V to 180V dynamic range at the High Voltage Digital Analog Converters. Most of the power is dissipated at each 96 Channel HV-DAC. There are 11 of them, ranging from 49.835°C for the coolest one (HV3) and 60.588°C for the hottest one (HV11). These tests were performed without the heat sink for the HV-DACs, which are currently being manufactured.

The right column of Table shows the voltage response v/s time for **incremental voltage commands of 100mV, 10V and 50V**, which takes 0.03ms, 0.3ms and 1ms respectively with less than 1% overshoot, and less 0.1ms settling time. The DM Controller meets the design requirements in terms of dynamic range of 180V and 16bit resolution, which it is equivalent to 2.7mV for the LSB. The noise level is within 10mV RMS, meeting the14bits accuracy requirement.

### 3.3 Optical testing

We validated the functionality and performance of the DM controller performing optical tests using a ZYGO interferometer. We installed a magnification system in front of the ZYGO in order to resolve maximum stroke deltas between adjacent actuators, which have a 300μm pitch. Fig. 10 shows the optical layout used for testing. A spherical lens was installed at the exit of the ZYGO followed by a flat fold mirror and a collimation lens that provides a 8:1 magnification. The interferometer operates with a 632.8nm wavelength laser.

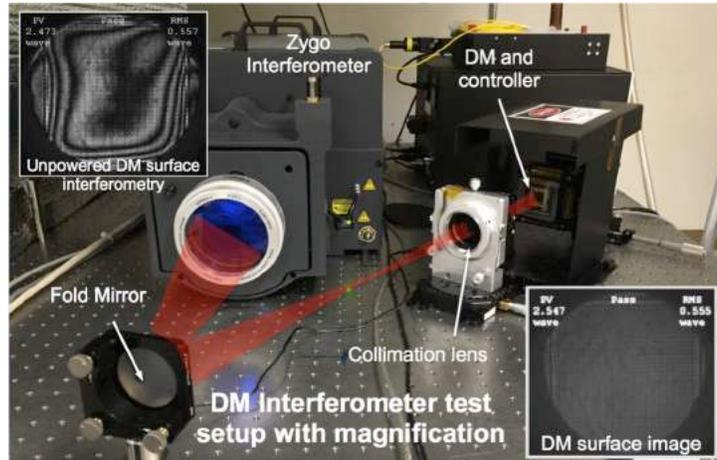

We mounted the system on the air-floating bench available at the ACE lab, therefore solving the vibration issues. However turbulence from the air conditioning system became the main source of noise during the measurements. We characterized the noise by taking 10 measurements with the DM powered, and applying a uniform voltage of 50V to all the actuators. The result is an average noise over the 10 samples of 58.4nm PV, and 7nm RMS.

Fig. 10: Laboratory test setup

For initial testing we used a damaged Kilo DM shown in Fig. 4b that has some functional areas. This device allowed us to safely perform initial full stroke test and the correct operation of the controller. Afterwards, we mounted a fully functional Kilo DM for the optical testing. The first step was to measure the unpowered surface shape (Fig. 11 top), which exhibits slightly more than 3 waves PV, and it was saved as a reference. For the following measurements, the reference was digitally subtracted from each measurement. Fig.11 at the bottom shows the residuals after measuring again with no voltage change and subtraction on. The result is 56.3nm PV and 4.4nm RMS, which is consistent with noise estimation of 58.4nm PV and 6.9nm RMS based on 10 consecutive measurements of the DM surface. The main component of the noise is the turbulence caused by the air conditioning system and the interferometer measurement error.

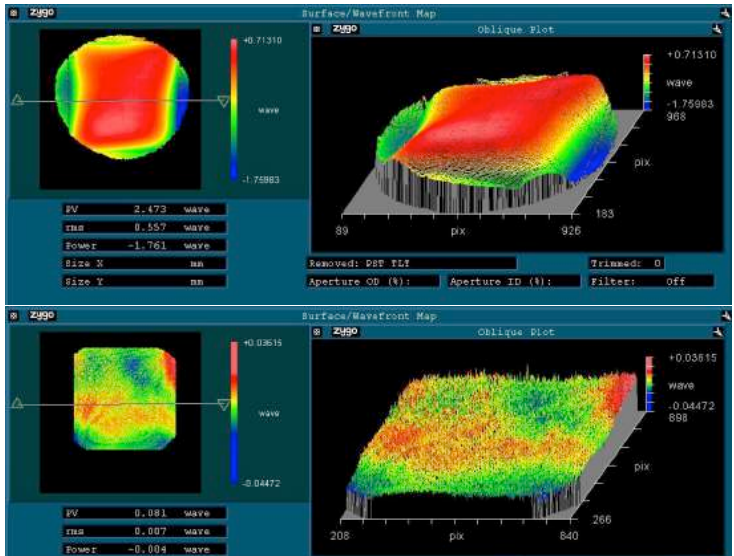

Fig. 11: Top. DM surface without any voltage applied. Bottom, DM surface after subtract the DM shape measured on top.

We continued the characterization with a reproducibility test for which an area of the mirror of 5x10 actuators was cycled between 50V and 120V ten times. We took a measurement after each voltage change to obtain the reproducibility of the system. We found an error of 39 +/- 58.4nm PV and 4 +/-7nm RMS. The error measured is within the noise, so we can say that there were no measurable reproducibility errors. In the future we plan to perform this kind of characterization measuring the brightness of speckles in the coronagraph. This will allow us to assess the mirror motion with significantly less noise. Then, we performed the first functional test involving all the actuators in the mirror. We applied 160V PV sinusoidal waves on the mirror surface of 3, 6 and 9 cycles per aperture in three different orientations. The 6 cycles per aperture in vertical and horizontal configuration and the vertical 9 cycles per aperture are shown in Fig. 12. The irregularities shown on the mirror surface corresponds to the channels controlled by one of the HV-DAC which was damaged during manufacturing because a soldering problem. The board manufacturer reported this problem and at the moment of writing the chip is being replaced. The other 10 HV-DAC show a 100% functional actuators.

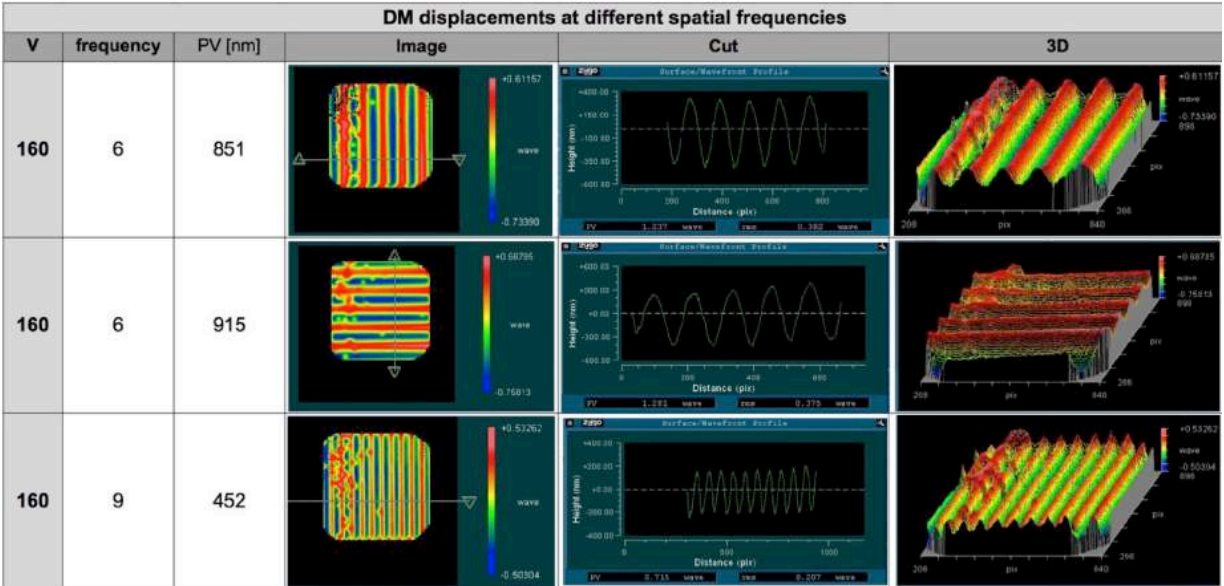

Fig. 12: DM Voltage response analysis. A region of 5x10 actuators was used to study the DM response. A uniform 50V was applied to the whole DM to provide a bias. 5V steps were applied from 5 to 160V. The error bars correspond to +/-29.5 nm, which is the interferometer measurement error for this configuration.

The next test was the DM response versus voltage applied. DM response. We selected a test region of 5x10 actuator to study the DM response. A uniform 50V was applied to the DM to provide a bias. Voltages from 5 to 160V, in 5V steps, were applied to the test region and the PV distance between the mirror surface and the test region was measured. Electrostatic

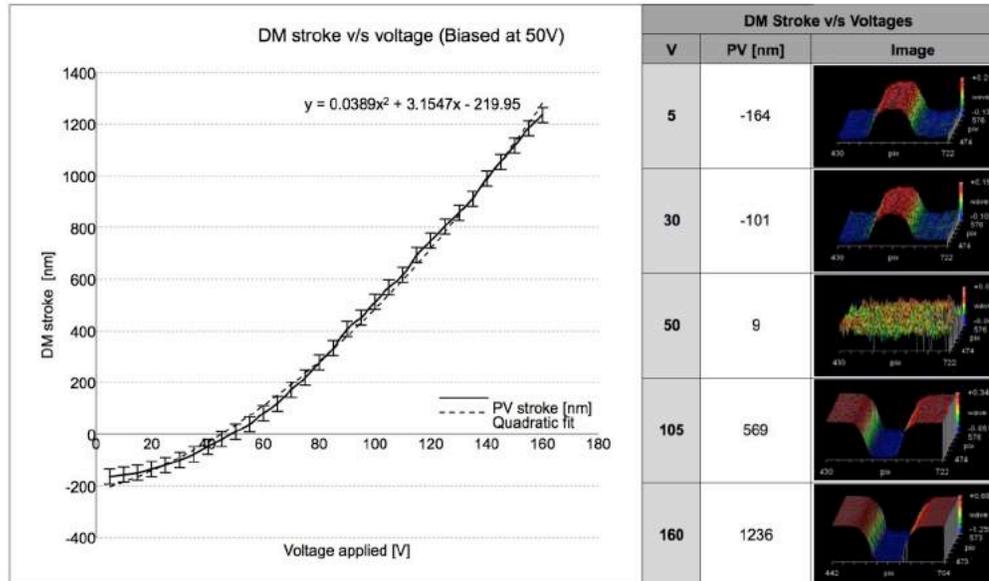

Fig. 8: DM Voltage response analysis. A region of 5x10 actuators was used to study the DM response. A uniform 50V was applied to the whole DM to provide a bias. 5V steps were applied from 5 to 160V. The error bars correspond to +/-29.5 nm, which is the interferometer measurement error for this configuration.

MEMS DMs such as the Kilo DM exhibit a nonlinear surface motion response to voltage applied. The analytical expression of the DM surface displacement in first approximation could be presented with a quadratic function. For this experiment we fitted the data with a quadratic function with reasonable results. Analysis of the DM response is beyond the scope of this paper and therefore no further investigation was done. The error bars correspond to +/-29.5 nm, which is the measurement error for this configuration.

## 4. CONCLUSIONS

We have developed a miniaturized wavefront control system that utilizes a BMC Kilo DM and a novel DM controller. The system architecture is based on an array of 96 channel High-Voltage Digital Analog Converters ASICs. Eleven ASICs are arranged in parallel to control up to 1056 channels. A single microprocessor receives a FITS file with the voltages for the actuators and sends the commands to all ASICs simultaneously. This architecture allows easy scalability to larger actuator counts.

The miniaturized DM controller has the same, or better, performance than the previous state of the art. It has 16bit discretization, with 14bit effective control over the noise, over a maximum dynamic range of 220V. The first prototype developed was running at a maximum of 180V to avoid DM over voltage in case there is problem with the controller. The system consumes 6W, fits in 9.5x9.5x60 mm, which fits in a 1.5U cubesat format and it weights less than 0.5kg.

The system has been designed to evolve into a space capable version without major modifications. The vibration and thermal management is space compatible. Most of the cables are isolated with Kapton to avoid outgassing. The main improvements to evolve to a space capable version is to add radiation shielding, coat all the electronics and PCBs to avoid outgassing and ensure allow mechanical strain relief for differential expansion of the PCB and the boards. We have envisioned solutions for all this challenges. Moreover, the current unit can easily evolve into a cryo-vacuum compatible version performing similar upgrades to the space capable version.

We hope that this controller will enable a wide range of new applications for MEMS DMs in ground and space, making small exoplanet imaging missions possible, as well as more compact and better performance future ground-based adaptive optics systems. We also hope that a more versatile controller will increase the demand and significantly reduce costs of controllers and DM given economy of scale savings.

## 5. ACKNOWLEDGEMENTS

We would like to acknowledge the NASA Ames Center Innovation Fund grant awarded to this project for fiscal year 2016 that enabled this development. Also we would like to thanks Boston Micro machines Corporation for their help and assistance to develop this device.